\documentstyle[preprint,prb,aps,epsf,floats,amssymb]{revtex}

\begin{document}


\draft

\title{Two interacting particles in a random potential:\\The random
  matrix model revisited }

\author{Thomas Vojta, Rudolf A. R\"omer and Michael Schreiber} 
\address{Institut f\"{u}r Physik, Technische
  Universit\"{a}t Chemnitz, D-09107 Chemnitz, Germany}

\date{$Revision: 2.1 $; compiled \today}

\maketitle

\begin{abstract}
  We reinvestigate the validity of mapping the problem of two onsite
  interacting particles in a random potential onto an effective random
  matrix model. To this end we first study numerically how the
  non-interacting basis is coupled by the interaction. Our results
  indicate that the {\em typical} coupling matrix element decreases
  significantly faster with increasing single-particle localization
  length than is assumed in the random matrix model.  We further show
  that even for models where the dependency of the coupling matrix
  element on the single-particle localization length is correctly
  described by the corresponding random matrix model its predictions
  for the localization length can be qualitatively incorrect.  These
  results indicate that the mapping of an interacting random system
  onto an effective random matrix model is potentially dangerous. We
  also discuss how Imry's block-scaling picture for two interacting
  particles is influenced by the above arguments.
\end{abstract}


\maketitle



%
%

\section{Introduction}

The interplay of disorder and many-body interactions in electronic
systems has been studied intensively within the last two
decades.\cite{review} For non-interacting electrons, the highly
successful ``scaling hypothesis of localization'' was put forward in
1979 by Abrahams {\em et al.},\cite{go4} but the role played by
many-particle interactions is much less understood and still no
entirely consistent picture exists.\cite{review} The recent discovery
of a metal-insulator transition in certain two-dimensional electron
gases at zero magnetic field \cite{kravch} has renewed the interest in
this problem, since in the samples considered the electron interaction
is estimated to be much larger than the Fermi energy. \cite{kravch}
Thus the observed transition may be due to an interaction-driven
enhancement of the conductivity.

The simplest version of the interacting disordered particle problem is
perhaps the case of just two interacting particles (TIP) in a random
potential in one dimension (1D).  For a Hubbard on-site interaction
this problem has recently also attracted a lot of attention after
Shepelyansky \cite{shep,shep2} argued that attractive as well as
repulsive interactions between the two particles (bosons or fermions)
lead to the formation of particle pairs whose localization length
$\lambda_2$ is much larger than the single-particle (SP) localization
length $\lambda_1$.\cite{numtmm} Based on a mapping of the TIP
Hamiltonian onto an effective random matrix model (RMM) he predicted
\begin{equation}
  \lambda_2 \sim (U/V)^2 \lambda_1^2
  \label{eq-shep}
\end{equation}
at two-particle energy $E=0$, with $V$ the nearest-neighbor transfer
matrix element and $U$ the Hubbard interaction strength.  Shortly
afterwards, Imry \cite{imry} used a Thouless-type block-scaling
picture (BSP) in support of this.  The most surprising aspect of Eq.\ 
(\ref{eq-shep}) is the fact that in the limit of weak disorder the
ratio $\lambda_2/\lambda_1$ diverges.  Thus, in the limit of weak
disorder the particle pair can travel infinitely further than a SP.
This should be contrasted with renormalization group studies of the 1D
Hubbard model at finite particle density which indicate that a
repulsive onsite interaction leads to a strongly localized ground
state.\cite{giamarchi}

Subsequent analytical investigations further explored the mapped TIP
problem as an RMM problem.\cite{jacquod1,weinmann,frahm2,frahm3} 
Direct numerical approaches to the TIP problem have been based on the
time evolution of wave packets, \cite{shep} transfer matrix methods
(TMM),\cite{frahm} Green function approaches,\cite{oppen,kim} or exact
diagonalization \cite{wein2}.  In these investigations usually an
enhancement of $\lambda_2$ compared to $\lambda_1$ has been found but
the quantitative results differ both from the analytical prediction in
Eq.\ (\ref{eq-shep}), and from each other. Furthermore, a check of the
functional dependence of $\lambda_2$ on $\lambda_1$ is numerically
very expensive since it requires very large system sizes.
Following the approach of Ref.\ \onlinecite{frahm}, two of us studied
the TIP problem by a different TMM \cite{rar1} and found that (i) the
enhancement $\lambda_2/\lambda_1$ decreases with increasing system
size $M$, (ii) the behavior of $\lambda_2$ for $U=0$ is equal to
$\lambda_1$ in the limit $M\rightarrow\infty$ only, and (iii) the
enhancement $\lambda_2/\lambda_1$ also vanishes completely in this
limit. Therefore it was concluded \cite{rar1} that the TMM applied to
the TIP problem in 1D measures an enhancement of the localization
length which is entirely due to the finiteness of the systems
considered.

In this paper we return the attention to the original mapping
\cite{shep} of the TIP problem onto an effective RMM. We argue that
the mapping as in Ref.\ \onlinecite{shep} is potentially dangerous
since (A) it overestimates the typical coupling matrix element and (B)
it neglects phase correlations which we believe to be essential,
because it is known that interference effects are responsible for
Anderson localization to begin with.
In order to establish that the mapping procedure \cite{shep} can lead
to incorrect results we first numerically investigate the
interaction-induced coupling matrix elements between the
non-interacting basis states for various values of the SP localization
length. We find that the typical coupling matrix element decreases
significantly faster with increasing SP localization length than
assumed in Ref.\ \onlinecite{shep}.  This alone would lead to a
significantly smaller increase, if any, of the TIP localization length
than in Eq.\ (\ref{eq-shep}). We further show that even if the RMM
correctly described the dependency of the coupling matrix element on
the SP localization length, its results for the TIP localization
length cannot be trusted. To this end we present two simple physical
examples, namely Anderson models with additional perturbing random
potentials for which the RMM mapping yields the same enhancement of
the localization length as for the TIP problem.  However, for our
examples this enhancement is obviously incorrect. We also show that
analogous problems exist for the BSP.\cite{imry} We argue that the
failure of the RMM approach in our toy models is caused by neglecting
the correlations between the coupling matrix elements.  This has
already been made responsible \cite{frahm,oppen} for quantitative
differences between Eq.\ (\ref{eq-shep}) and numerical results for the
TIP problem.  We show, however, that neglecting the correlations not
only changes the quantitative predictions of the theory but can lead
to qualitatively incorrect results.

The paper is organized as follows. In section \ref{sec-rmm} we briefly
summarize the RMM approach to the TIP problem. In section
\ref{sec-num} we present our numerical results for the TIP coupling
matrix elements and their dependence on the SP localization length.
The failure of the RMM approach to correctly predict the localization
length of two toy models is discussed in section \ref{sec-toy} while
section \ref{sec-bsptoy} shows the failure of the BSP for these toy
models. We discuss the relevance of our results for the original TIP
problem and conclude in section \ref{sec-con}.

%
%

\section{The random matrix model approach}
\label{sec-rmm}

Let us start by recalling the basic steps of the RMM approach
\cite{shep} to TIP in a random potential. The relevant energy scales
are chosen such that the SP band width $4 V$ is larger than the
(uniform) spread of the disorder $W$ which in turn is supposed to be
larger than the interaction strength $U$. The basic idea is to
represent the TIP Hamiltonian in the eigenbasis of the non-interacting
problem and then to replace the full Hamiltonian by a suitably chosen
random matrix.

The (non-interacting) SP eigenstates are approximately described by
\begin{equation}
  \psi_n (x) \sim {1 \over \sqrt\lambda_1} \exp \left [ - \frac
    {|x-x_n|} {\lambda_1}+i \theta_n(x) \right ],
  \label{eq-psi1}
\end{equation}
where $x_n$ is the localization center of the $n$th eigenstate and
$\theta_n(x)$ is a phase which appears to be random but contains all
the information about interferences necessary for Anderson
localization. In the absence of interactions and neglecting symmetry
considerations the two-particle eigenstates are just products of two
SP eigenstates,
\begin{equation}
  \psi_{nm} (x,y) \sim {1 \over \lambda_1}
  \exp \left [ - \frac {|x-x_n|} {\lambda_1} - \frac{|y-y_m|} {\lambda_1} 
  +i \theta_n(x)+i \theta_m(y) \right],
 \label{eq-psi2}
\end{equation}
where $x$ and $y$ are the coordinates of the first and second
particle, respectively. Switching on the Hubbard interaction $U(x,y)=
U\delta_{xy}$ between the two particles induces transitions between
the eigenstates $\psi_{nm}$ of the non-interacting problem.  To
estimate the transition rates it is first noted that the matrix
element $\langle \psi_{nm}|U|\psi_{n'm'} \rangle$ is exponentially
small for $|x_n - y_m| > \lambda_1$ or $|x_{n'} - y_{m'}| > \lambda_1$
or $|x_n - x_{n'}| > \lambda_1$ or $|y_m- y_{m'}| > \lambda_1$. Thus,
the interaction couples each of the two-particle states
(\ref{eq-psi2}) close to the diagonal in the 2D configuration space
($|x_n - y_m| < \lambda_1$) to ${\cal O} (\lambda_1^2)$ other such
states. The interaction matrix element is then the sum of $\lambda_1$
contributions each with magnitude $U \lambda_1^{-2}$ and approximately
random phases. Neglecting possible correlations among these
contributions, Shepelyansky found the magnitude $u$ of the matrix
element
\begin{equation}
  u_{nmn'm'} = 
  \langle \psi_{nm}|U|\psi_{n'm'} \rangle 
  \sim U \lambda_1^{-3/2} ,
  \label{eq-matrix}
\end{equation}
independent of the interaction being attractive, repulsive or even
random. Eq.\ (\ref{eq-matrix}) is one of the essential ingredients of
the RMM. In section \ref{sec-num} we will present numerical data in
order to check its validity. We remark that the validity of Eq.\ 
(\ref{eq-matrix}) has recently been questioned in Ref.\ \onlinecite{PS}
where the authors have computed a different estimate taking into
account the nearly Bloch-like structure of the eigenstates for small
$W$.

Shepelyansky \cite{shep,shep2} now replaced the full TIP Hamiltonian
by an effective RMM for those of the two-particle states that are
coupled by the interaction.  Thus the Hamiltonian matrix becomes a
banded matrix whose elements are independent Gaussian random numbers
with zero mean.  The diagonal elements are drawn from a distribution
of width $V$, because for small disorder $W$ the nearest-neighbor
transfer $V$ determines the band width of the SP states. The
distribution of the off-diagonal elements has width
$|U|\lambda_1^{-3/2}$ within a band of width $\lambda_1$.

In order to obtain results for the localization properties of such an
RMM one has to distinguish different regimes, depending on the
strength of the interaction. If the interaction is strong enough to
couple many non-interacting eigenstates, {\em i.e.}, the inverse
lifetime $\Gamma$ of a non-interacting state is large compared to the
level spacing of the coupled states, Fermi's golden rule can be
applied.  This regime was investigated in Ref.\ \onlinecite{shep} and
also gives the largest enhancement of $\lambda_2$ compared to
$\lambda_1$. We note that in this regime the level-spacing
distribution of the non-interacting system cannot play a significant
role since the interactions couple a large number of levels and lead
to a decay into a quasi-continuum of final states.  In the opposite
limit, {\em i.e.}, if the interaction couples only few non-interacting
eigenstates, Fermi's golden rule cannot be applied.  Instead, one
finds Rabi oscillations between the few coupled states.\cite{weinmann}
We note that in this regime the level spacing distribution of the
non-interacting states becomes important. In the following we will
only consider the golden rule regime.

The localization length of the effective RMM can be determined by
several equivalent methods. Here we follow Shepelyansky:\cite{shep2}
Calculating the decay rate $\Gamma$ of a non-interacting eigenstate by
means of Fermi's golden rule gives $\Gamma \sim U^2/\lambda_1 V$.
Since the typical hopping distance is of the order of $\lambda_1$ the
diffusion constant is $D \sim U^2 \lambda_1 /V$.  Within a time $\tau$
the particle pair visits $N \sim U \lambda_1^{3/2} V^{-1/2}
\tau^{1/2}$ states.  Diffusion stops when the level spacing of the
visited states is of the order of the frequency resolution $1/\tau$.
This determines the cut-off time $\tau^*$ and the corresponding
pair-localization length is obtained as $\lambda_2 \sim \sqrt{ D
  \tau^*} \sim (U/V)^2 \lambda_1^2$ in agreement with Eq.\ 
(\ref{eq-shep}). Applicability of Fermi's golden rule requires $\Gamma
\gg V/\lambda_1^2$ which is equivalent to $U^2 \lambda_1 /V^2 \gg 1$.
This is exactly the condition for an enhancement of $\lambda_2$
compared to $\lambda_1$.
 
Let us recapitulate: The mapping of the TIP problem onto the RMM
described above relies on two assumptions: (A) the non-interacting
wavefunctions can be described by a decaying amplitude with finite
localization length and a random phase which leads to the
$U/\lambda_1^{3/2}$ behavior in Eq.\ (\ref{eq-matrix}) and (B) any
correlations between the matrix elements in the Hamiltonian can be
neglected.  In the next two sections we will closer analyze the
validity of these two assumptions.

%
%

\section{Numerical results for the matrix elements}
\label{sec-num}

In this section we present results for the interaction-induced
coupling matrix elements in order to check whether they follow the
$\lambda_1^{-3/2}$ power law (\ref{eq-matrix}) as assumed in Ref.\ 
\onlinecite{shep}. Since $\lambda_1$ deviates from the simple
power-law prediction \cite{numtmm} $\lambda_1 \approx 104/W^2$ at
$E=0$ already for $\lambda_1 \lesssim 4$ ($W\gtrsim 5$), we have first
computed $\lambda_1$ by TMM \cite{numtmm} in 1D with $0.1\%$ accuracy
for all $W\geq 0.3$ ($\lambda_1\approx 1156$). We next exactly
diagonalize the SP Hamiltonian and obtain the eigenstates.  We then
compute the ``center-of-mass'' (CM) of these eigenstates as $x_n =
\sum_{x} x |\psi_{n}(x)| / \sum_{x} |\psi_{n}(x)|$. For hard wall
boundary conditions, we have checked that using this definition of the
CM we can reproduce the disorder dependence of $\lambda_1$ from the
decay of the SP wave function $\psi_n$ via $1/\lambda_1 = -
\lim_{|x-x_n|\rightarrow\infty} \ln |\psi_n(x)| / |x-x_n|$ to within
$10\%$ up to $\lambda_1 = 104$ ($W=1$) for $50$ samples of length
$M=1200$. For periodic boundary conditions, we use a suitably
generalized definition for the CM. We next calculate the matrix
elements $\langle \psi_{nm}|U|\psi_{n'm'} \rangle$ for all states with
appropriate CM, {\em i.e.}, $|x_n - y_m| \leq \lambda_1$, $|x_{n'} -
y_{m'}| \leq \lambda_1$, $|x_n - x_{n'}| \leq \lambda_1$ and $|y_m-
y_{m'}| \leq \lambda_1$. Since the interaction strength $U$ appears
only as a multiplicative prefactor in the matrix elements, we choose
$U=1$ in all of what follows. We emphasize that the bottleneck in such
a computation is not the system size $M$, but rather the exponentially
growing number of overlapping matrix elements for increasing
$\lambda_1$.

\begin{figure} 
   \epsfxsize=0.8\columnwidth
   \centerline{\epsffile{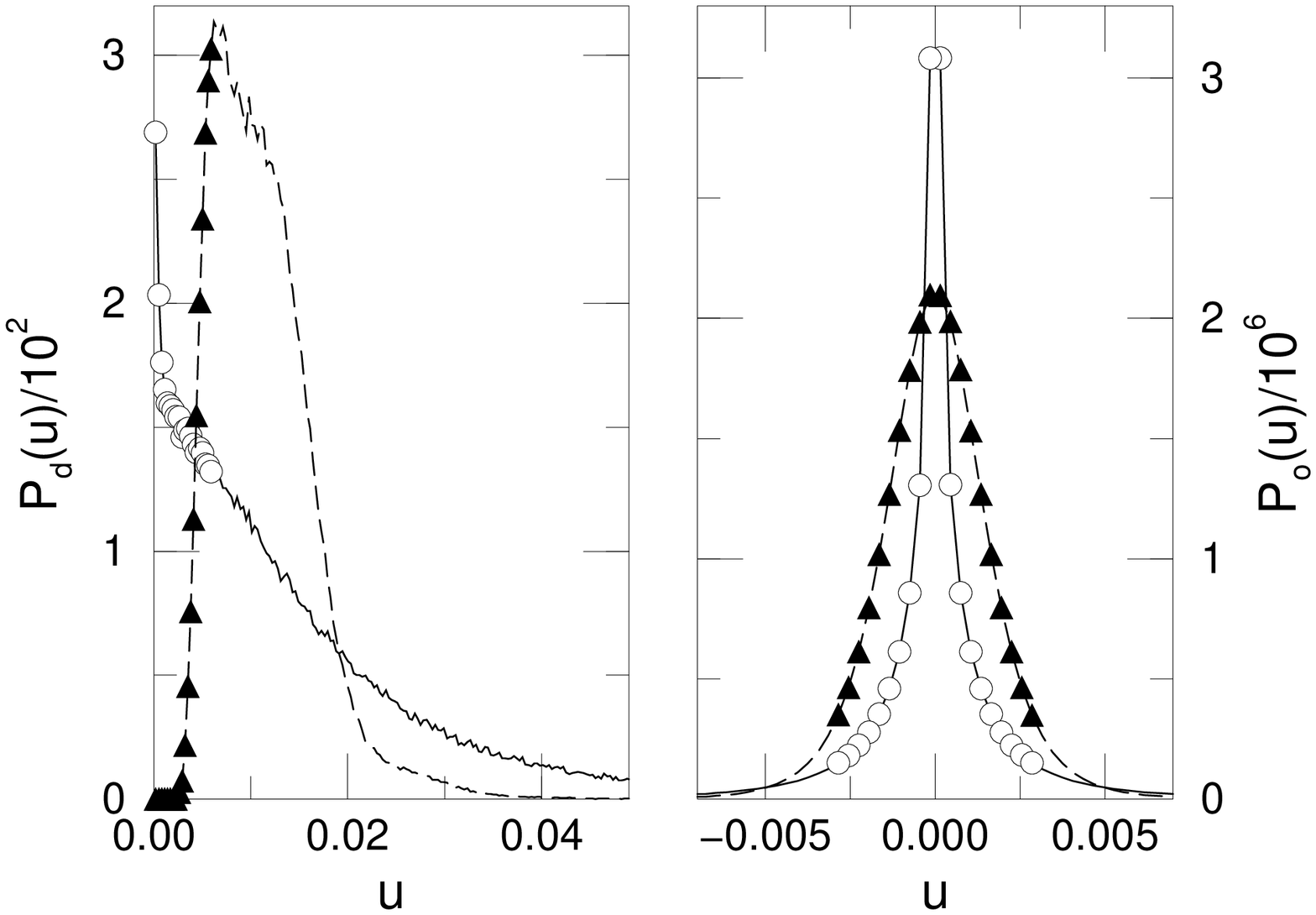}}
   \caption{
     Unnormalized distribution $P_{\text{d/o}}(u)$ of the diagonal
     (left panel) and off-diagonal (right panel) coupling matrix
     elements $u$ with bin width $\Delta= 0.0003$ for $\lambda_1=26$
     ($W=2$) and $M=200$.  Circles and solid lines indicate TIP data,
     triangles and dashed lines indicate matrix elements computed
     using Eq.\ (\protect\ref{eq-psi2}). The symbols mark the data for
     the 20 smallest $|u|$.}
   \label{fig-tipme-bin}
\end{figure}

\begin{figure} 
   \epsfxsize=0.8\columnwidth
   \centerline{\epsffile{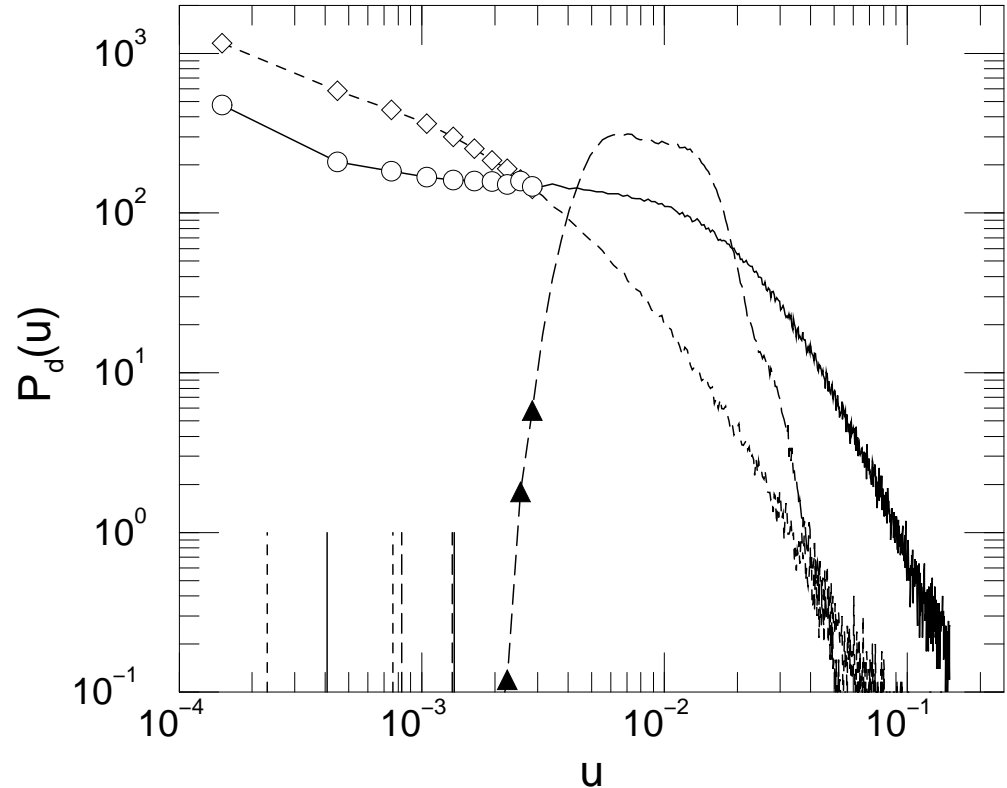}}
   \caption{
     Double-logarithmic plot of the unnormalized distribution
     $P_{\text{d}}(u)$ of the diagonal coupling matrix elements $u$
     for $\lambda_1=26$ ($W=2$) and $M=200$ as in Fig.\ 
     \protect\ref{fig-tipme-bin}. Solid, short-dashed, and long-dashed
     lines correspond to the TIP problem, the TIP problem with random
     interaction, and Eq.\ (\protect\ref{eq-psi2}), respectively.
     Circles, diamonds and triangles mark the data for the $10$
     smallest $u$ in each case, only three of which are larger than
     $10^{-1}$ for the triangles. The vertical lines on the $u$-axis
     indicate $u_{\text{abs}}$ (right) and $u_{\text{typ}}$ (left) as
     computed from the total distribution $P(u)$.}
   \label{fig-tipme-bin-DD}
\end{figure}

\begin{figure} 
   \epsfxsize=0.8\columnwidth
   \centerline{\epsffile{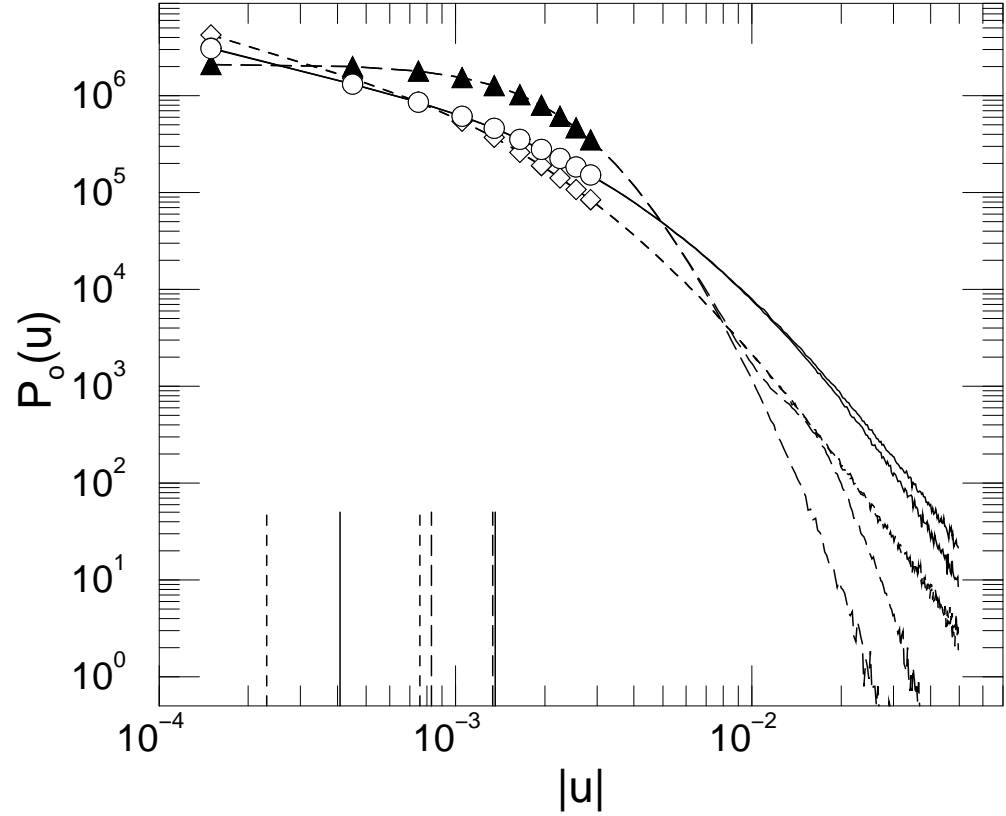}}
   \caption{
     Double-logarithmic plot of the unnormalized distribution
     $P_{\text{o}}(u)$ of the off-diagonal coupling matrix elements
     $u$ for $\lambda_1=26$ ($W=2$) and $M=200$ as in Fig.\ 
     \protect\ref{fig-tipme-bin}. Solid, short-dashed, and long-dashed
     lines correspond to the TIP problem, the TIP problem with random
     interaction, and Eq.\ (\protect\ref{eq-psi2}), respectively.
     Circles, diamonds and triangles mark the data for the $10$
     smallest $|u|$ in each case. The vertical lines on the $u$-axis
     indicate $u_{\text{abs}}$ (right) and $u_{\text{typ}}$ (left) as
     computed from the total distribution $P(u)$.}
   \label{fig-tipme-bin-OD}
\end{figure}

In Fig.\ \ref{fig-tipme-bin} we show the unnormalized probability
distributions $P_{\text{d/o}}(u)$ of diagonal and off-diagonal
coupling matrix elements.  $P_{\text{d/o}}(u)$ was computed at
disorder $W=2$ where the enhancement of $\lambda_2$ with respect to
$\lambda_1$ is expected to be large.\cite{frahm,oppen,kim} We have
averaged over $50$ different disorder configurations for $M=200$.  For
a more detailed inspection we plot the data on a doubly logarithmic
scale in Figs.\ \ref{fig-tipme-bin-DD} and \ref{fig-tipme-bin-OD}.  As
already discussed before \cite{frahm,oppen,evangelou} we note that (i)
the diagonal elements are non-negative, (ii) $P_{\text{o}}(u)$ is
symmetric around $u=0$. The deviation from symmetry for $|u| \gtrsim
0.02$, {\em i.e.}, $P_{\text{o}}(|u|) > P_{\text{o}}(-|u|)$, is most
likely due to the finite size of the samples.  More importantly, (iii)
$P_{\text{o}}(u)$ is strongly non-Gaussian. We remark that a fit to a
Lorentzian distribution does also not describe the data.  (iv) apart
from a peak at $u \approx 0$, $P_{\text{d}}(u)$ is approximately
Gaussian, (v) $P_{\text{o}}(u)$ and $P_{\text{d}}(u)$ have rather long
tails, (vi) the total distribution of matrix elements $P(u)$ is
dominated by $P_{\text{o}}(u)$ (as in any matrix) and thus $P(u)$ is
strongly non-Gaussian with long tails.

For such a non-Gaussian $P(u)$ the average of the absolute matrix
elements $u_{\text{abs}}= \langle |u| \rangle$, with $\langle \cdot
\rangle$ denoting the average over $u$ according to $P(u)$, is
strongly influenced by rare events in the tails of the distribution.
This is even more so when using the mean-square value $\sqrt{\langle
  u^2 \rangle}$. However, in the physical problem considered here
these rare large couplings lead to oscillations of the system between
the corresponding two TIP states but not to delocalization. The {\em
  typical} value $u_{\text{typ}}$ is thus better defined as the
logarithmic average $u_{\text{typ}}= \exp[\langle \log(|u|)\rangle]$.

\begin{figure} 
   \epsfxsize=0.8\columnwidth
   \centerline{\epsffile{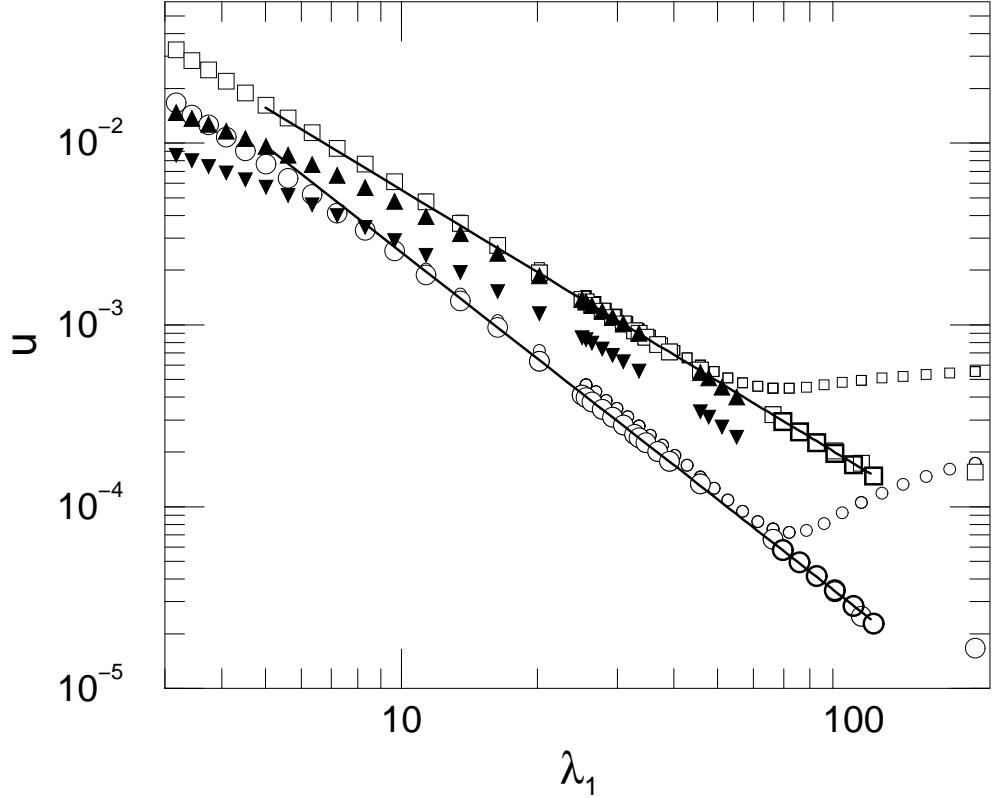}}
   \caption{
     Dependence of $u_{\text{abs}}$ ($\Box$, $\blacktriangle$) and
     $u_{\text{typ}}$ ($\circ$, $\blacktriangledown$) on $\lambda_1$
     for the TIP eigenstates (open symbols) and states chosen
     according to Eq.\ (\protect{\ref{eq-psi2}}) (filled symbols) for
     $M=200$. The small (bold) symbols indicate $u_{\text{abs}}$ and
     $u_{\text{typ}}$ for $M=100$ ($M=250$). The solid lines represent
     the power laws $u_{\text{abs}} \sim \lambda_1^{-1.5}$ and
     $u_{\text{typ}} \sim \lambda_1^{-1.95}$. }
   \label{fig-tipme-HW-l1}
\end{figure}

We have calculated both $u_{\text{abs}}$ and $u_{\text{typ}}$ for
different values of $W$ and $50$ samples for $M \leq 200$ and $30$
samples for $M=250$. As shown in Fig.\ \ref{fig-tipme-HW-l1}, the
dependence of $u_{\text{abs}}$ on $\lambda_1$ for $\lambda_1 > 5$
follows $u_{\text{abs}} \propto \lambda_1^{-\alpha}$. A fit for $20
\leq \lambda_1 \leq 111$ yields $\alpha = {-1.5 \pm 0.1}$ as
predicted in Ref.\ \onlinecite{shep}.  However, the typical TIP matrix
element $u_{\text{typ}}$ decreases much faster with increasing
$\lambda_1$.  Fitting the $u_{\text{typ}}$ data to a power law for $20
\leq \lambda_1 \leq 111$, we obtain $\alpha=1.95 \pm 0.10$.
Furthermore, for the largest $\lambda_1$ the numerical data deviate
--- albeit weakly --- from the above power law showing a slight
downward curvature in the double-logarithmic representation. In fact,
if we only consider the data points from the large chains with
$M=250$, we already find $\alpha=2.02 \pm 0.10$. This indicates that
asymptotically the dependence is even stronger than
$\lambda_1^{-1.95}$.

\begin{figure} 
  \epsfxsize=0.8\columnwidth
  \centerline{\epsffile{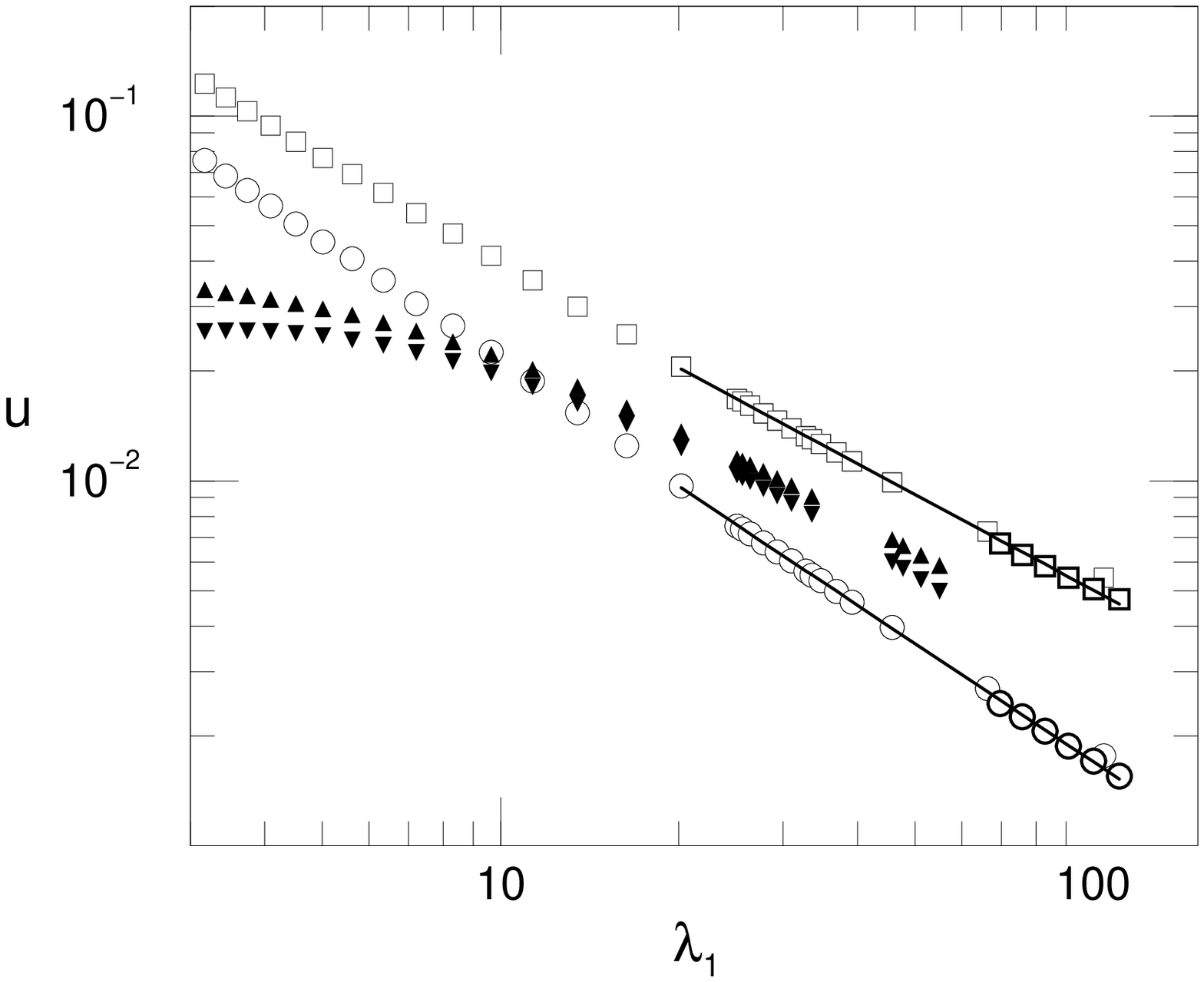}}
   \caption{
     Dependence of $u_{\text{d,abs}}$ ($\Box$, $\blacktriangle$) and
     $u_{\text{d,typ}}$ ($\circ$, $\blacktriangledown$), {\em i.e.},
     for diagonal matrix elements only, on $\lambda_1$ for the TIP
     eigenstates (open symbols) and states chosen according to Eq.\ 
     (\protect{\ref{eq-psi2}}) (filled symbols) for $M=200$. The bold
     symbols indicate $u_{\text{d,abs}}$ and $u_{\text{d,typ}}$ for
     $M=250$. The solid lines represent the power laws
     $u_{\text{d,abs}} \sim \lambda_1^{-0.9}$ and $u_{\text{d,typ}}
     \sim \lambda_1^{-1.0}$.}
   \label{fig-tipme-HW-DD-l1}
\end{figure}

In Fig.\ \ref{fig-tipme-HW-DD-l1}, we show $u_{\text{abs}}$ and
$u_{\text{typ}}$ for the diagonal matrix elements only. For $20
\leq \lambda_1 \leq 111$ the data can be fitted by
$u_{\text{d,abs}} \propto \lambda_1^{-0.9 \pm 0.1}$ and
$u_{\text{d,typ}}\propto \lambda_1^{-1.0 \pm 0.1}$.  Thus as expected
$u_{\text{d,abs}}$ and $u_{\text{d,typ}}$ behave similarly since
$P_{\text{d}}(u)$ may be approximated by a Gaussian distribution.
Furthermore, $\alpha=1$ is in agreement with Eq.\ (21) of Ref.\ 
\onlinecite{PS}.

Repeating the RMM calculation of section \ref{sec-rmm} with a
dependence $u_{\text{typ}} \sim U\lambda_1^{-\alpha}$ instead of Eq.
(\ref{eq-matrix}), we obtain
\begin{equation}
  \lambda_2 \sim (U/V)^2 \lambda_1^{-2\alpha+5}~.
\end{equation}
If we use now use $\alpha=1.95 \pm 0.10$ we find $\lambda_2 \sim
\lambda_1^{1.1 \pm 0.2}$.  However, as discussed above the true
asymptotic dependence of $u_{\text{typ}}$ on $\lambda_1$ is likely to
be even stronger than $\lambda_1^{-1.95}$ which in turn results in an
even weaker enhancement of $\lambda_2$. We emphasize that the
enhancement predicted by Shepelyansky \cite{shep,shep2} will vanish
for $\alpha = 2$ in the limit $\lambda_1 \to \infty$. A value of
$\alpha > 2$ will in fact correspond to even stronger localization of
the TIP.

In order to further explore the validity of assumption (A) we compute
$P_{\text{d/o}}(u)$ for a site-dependent random onsite interaction
$U(x) \in [-U,+U]$, averaging as before over $50$ samples.  If
assumption (A) is correct, the resulting distribution of the coupling
matrix elements should qualitatively be similar to the one obtained
for the original TIP problem. But as shown in Figs.\ 
\ref{fig-tipme-bin-DD} and \ref{fig-tipme-bin-OD}, we find that the
randomness of the interaction already leads to a significant decrease
of the long-range nature of $P(u)$. {\em E.g.}, at $|u|=0.02$, there
is a reduction in $P_{\text{d/o}}(0.02)$ by a factor of approximately
$10$ for diagonal and approximately $5$ for off-diagonal matrix
elements when compared to $P_{\text{d/o}}(0.02)$ of the original TIP
problem.
We further compute $P(u)$ for states with the same CM as previously,
but otherwise chosen according to Eq.\ (\ref{eq-psi2}) with
uncorrelated random phases and exponentially decaying envelope.  The
disorder averaging is again over $50$ samples. As shown in Fig.\ 
\ref{fig-tipme-bin}, $P_{\text{d}}(u)$ now has a maximum at finite
$u$. For these states $P_{\text{o}}(u)$ is well approximated by a
Gaussian just as expected by Shepelyansky.\cite{shep,shep2} The
double-logarithmic plot of Fig.\ \ref{fig-tipme-bin-OD} shows
deviations from the symmetry $P_{\text{o}}(u) = P_{\text{o}}(-u)$ for
$|u| \gtrsim 0.008$, {\em i.e.}, $P_{\text{o}}(|u|) >
P_{\text{o}}(-|u|)$. As for the TIP problem, we attribute this to the
finite size of the samples considered. Again, we note that when
compared to $P_{\text{o}}(u)$ for the TIP problem, the present
distribution of matrix elements decreases much faster and at
$|u|=0.02$ is about one order of magnitude smaller. Furthermore, for
$|u|>0.008$ the distribution $P_{\text{o}}(u)$ is also smaller than
that for the model with random interaction. Thus assumption (A)
clearly oversimplifies the problem and the neglect of phase
correlations leads to a wrong $P_{\text{d/o}}(u)$.
In Fig.\ \ref{fig-tipme-HW-l1}, we show $u_{\text{abs}}$ and
$u_{\text{typ}}$ for the artificial states of Eq.\ (\ref{eq-psi2}). In
complete agreement with our previous discussion, we find that for
$\lambda_1 > 20$ both the average and the typical matrix element vary
as $u \propto \lambda_1^{-1.4 \pm 0.1}$ compatible with $\alpha= 3/2$.

In Fig.\ \ref{fig-tipme-HW-l1}, we show also TIP data for chain
lengths $M=100$. We note that deviations due to the small system size
lead to a smaller slope for $u_{\text{abs}}$ and thus may give rise to
an apparent enhancement of $\alpha$. As can be seen in the figure,
this decreasing of the slope happens for $M=100$ already at $\lambda_1
\gtrsim 20$ ($W \lesssim 2.3$). A power-law fit for $30 \leq \lambda_1
\leq 57$ yields $u_{\text{abs}} \propto \lambda_1^{-1.39 \pm 0.10}$.
The finite-size deviations for $u_{\text{typ}}$ are different.  A
power-law fit for $7 \leq \lambda_1 \leq 30$ gives $u_{\text{typ}}
\propto \lambda_1^{-1.77 \pm 0.10}$ whereas for $30 \leq \lambda_1
\leq 57$ we find $u_{\text{typ}} \propto \lambda_1^{-2.06 \pm 0.10}$.
Thus first there is a decrease of $\alpha$ followed by a finite-size
increase of $\alpha$. For still larger $\lambda_1 \gg M/2$ the
finite-size deviations of $u_{\text{abs}}$ and $u_{\text{typ}}$ become
very large even resulting in a positive slope.  This finite-size
effect may be at least partially responsible for the enhancement
observed in Refs.\ \onlinecite{frahm} and \onlinecite{rar1} for this
value of $M$.

\begin{figure} 
   \epsfxsize=0.8\columnwidth
   \centerline{\epsffile{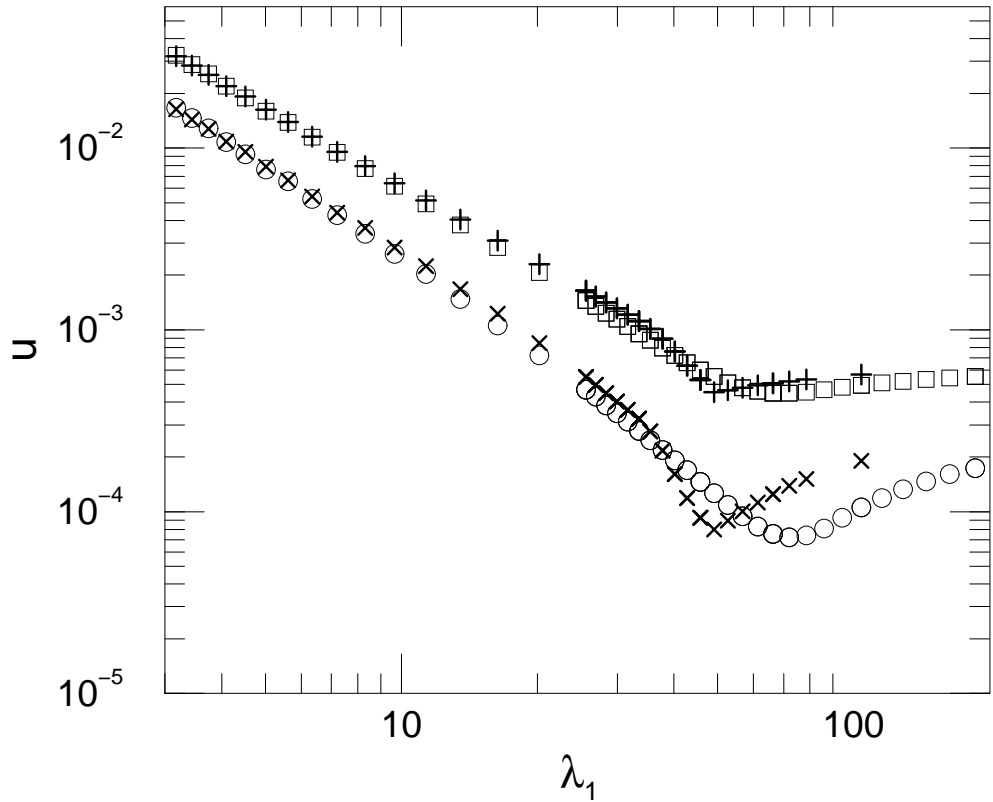}}
   \caption{
     Dependence of $u_{\text{abs}}$ ($\Box$, $+$) and $u_{\text{typ}}$
     ($\circ$, $\times$) on $\lambda_1$ for the TIP eigenstates for
     hard wall ($\Box$,$\circ$) and periodic ($+$, $\times$) boundary
     conditions and size $M=100$.  }
   \label{fig-tipme-HWPB-l1}
\end{figure}

In Fig.\ \ref{fig-tipme-HWPB-l1}, we show $u_{\text{abs}}$ and
$u_{\text{typ}}$ for $M=100$ with hard wall and periodic boundary
conditions. Up to $\lambda_1 \approx 10$, the data for both boundary
conditions agree quite well. For $10 \leq \lambda_1 \leq 25$, the
slope for the data with periodic boundaries is slightly smaller than
for the data with hard wall boundaries. Lastly, around $\lambda_1
\approx M/2$, the data for periodic boundaries shows a very fast
decrease of $u$. Thus the data for periodic boundaries is influenced
by the finite size of the sample already earlier than the data for
hard wall boundaries. Nevertheless, except for these finite size
effects, our results for both boundary conditions are similar and we
will restrict ourselves to the hard wall boundaries in the following.
We remark that most numerical studies of the TIP problem also use this
type of boundaries. \cite{frahm,oppen,kim,wein2,rar1}

%
%

\section{Failure of the RMM approach for toy models}
\label{sec-toy}

In this section we show that even an RMM which contains the correct
dependence of the coupling matrix elements on the SP localization
length may give qualitatively incorrect results. To this end we
consider two toy models, viz.\ Anderson models of localization with
additional perturbing random potentials. By a procedure analogous to
that of section \ref{sec-rmm} we map these models onto RMMs and then
show that these RMMs give erroneous enhancements of the localization
length.


\subsection{2D Anderson model with perturbation on a line}
\label{sec-2dam}

The first example is set up to lead to the same RMM as the TIP
problem. It consists of the usual 2D Anderson model of localization
perturbed by an additional weak random potential of strength $U$ at
the diagonal $x=y$ in real space. Since this increases the width of
the disorder distribution at the diagonal we expect the localization
length to decrease.  We now map onto an RMM as in Refs.\ 
\onlinecite{shep,shep2}.  As above, the eigenstates of the unperturbed
system are localized with a localization length $\lambda_1$ and
approximately given by
\begin{eqnarray}
  \psi_{n} (x,y) \sim {1 \over \lambda_1} \exp \left[ -\frac {|{\bf r}-{\bf
    r}_n|} {\lambda_1}+ i \theta_n({\bf r}) \right]
\end{eqnarray}
where ${\bf r}= (x,y)^T$ is the coordinate vector of the particle and
$\theta$ is again a phase which is assumed to be random.  The
Hamiltonian of this model differs from the TIP Hamiltonian in two
points: (i) the diagonal elements are independent random numbers
instead of being partially correlated as in the TIP problem and (ii)
the interaction potential $U(x,x) \in [-U,U]$ at each site of the
diagonal is random instead of having a definite sign and modulus $U$
as in the TIP problem. However, none of these points enters the
mapping procedure outlined above.  Thus, applying exactly the same
arguments as for the TIP problem in section \ref{sec-rmm} we find that the
perturbation couples each state close to the diagonal ($|x_n - y_n| <
\lambda_1$) to ${\cal O}(\lambda_1^2)$ other such states. The
interaction matrix element is again a sum of ${\cal O} (\lambda_1)$
terms of magnitude $U/\lambda_1^2$ and random phases giving a typical
value of $U\lambda_1^{-3/2}$.  Consequently, our toy model is mapped
onto exactly the same RMM as TIP in a random potential.

\begin{figure} 
   \epsfxsize=0.8\columnwidth
   \centerline{\epsffile{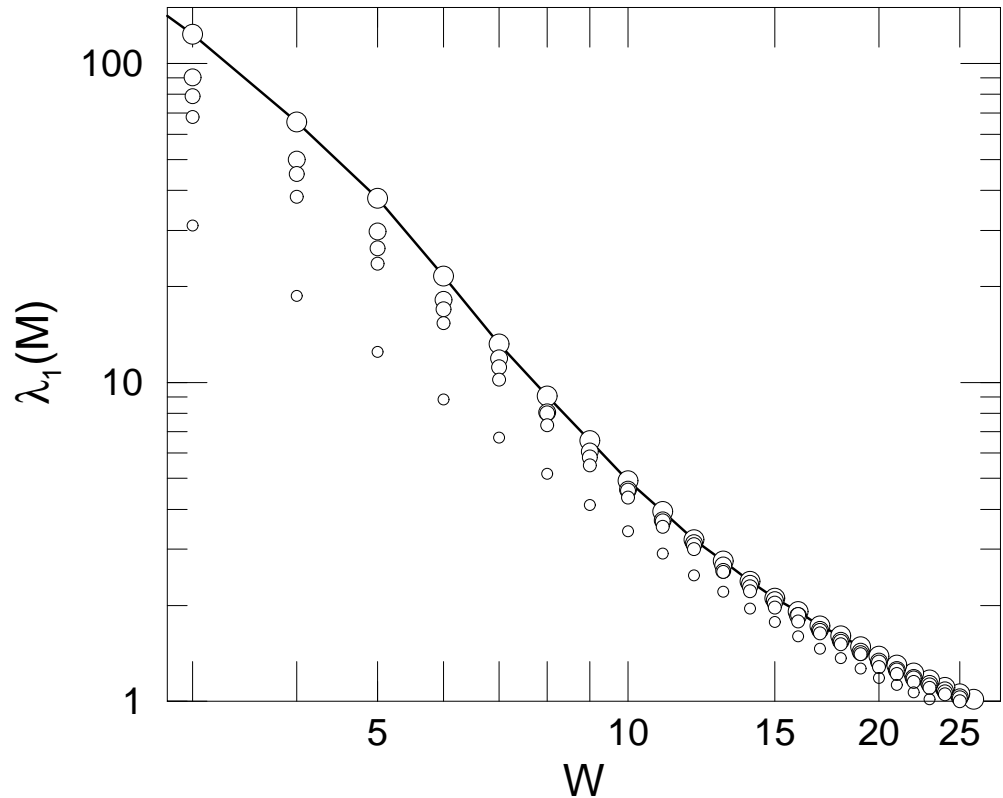}}
   \caption{
     Dependence of $\lambda_1(M)$ on disorder $W$ for the 2D Anderson
     model at $E=0$ for $M= 10, 25, 30, 35$ and $50$ indicated by
     increasing symbol size. We use the $M=50$ data, emphasized by the
     solid line, as finite-size estimate of $\lambda_1$. }
   \label{fig-d2me-l1w}
\end{figure}

As for the TIP case we now numerically check the relation between the
coupling matrix element and the SP localization length $\lambda_1$.
We first note that the disorder dependence of $\lambda_1$ in the 2D
Anderson model is no longer approximated by the simple power law cited
in section \ref{sec-num}.\cite{mkk} In fact, $\lambda_1$ is usually
much larger in the 2D case for the same value of $W$.  Thus we compute
estimates $\lambda_1(M)$ as a function of $W$ for quasi-1D strips
of finite strip width $M$ with $1\%$ accuracy by TMM. We remark that
due to the self-averaging \cite{mkk} of $1/\lambda_1(M)$ this is
equivalent to computing $\lambda_1(M)$ for many samples of $M \times
M$ disordered squares. In Fig.\ \ref{fig-d2me-l1w}, we show data of
$\lambda_1(M)$ as a function of $W$. We take $\lambda_1(50)$ to
compute the coupling matrix elements. Since $\lambda_1(50)$ is always
larger than for smaller system size, this choice only means that we
sum over a few additional but very small terms when computing $u$.
Next, we calculate both $u_{\text{abs}}$ and $u_{\text{typ}}$ for
different values of $W$ and various $M \times M$ squares. Disorder
averaging is over $20$ samples and we study $u_{\text{abs}}$ and
$u_{\text{typ}}$ as functions of $\lambda_1(M)$.  We
emphasize that instead of the well-known extrapolations of
$\lambda_1(M)$ to infinite system size by means of finite-size
scaling,\cite{mkk} we take the finite-size approximants $\lambda_1(M)$
on purpose, since we compute $\lambda_2$ also for comparable finite
sizes only.

\begin{figure} 
   \epsfxsize=0.8\columnwidth
   \centerline{\epsffile{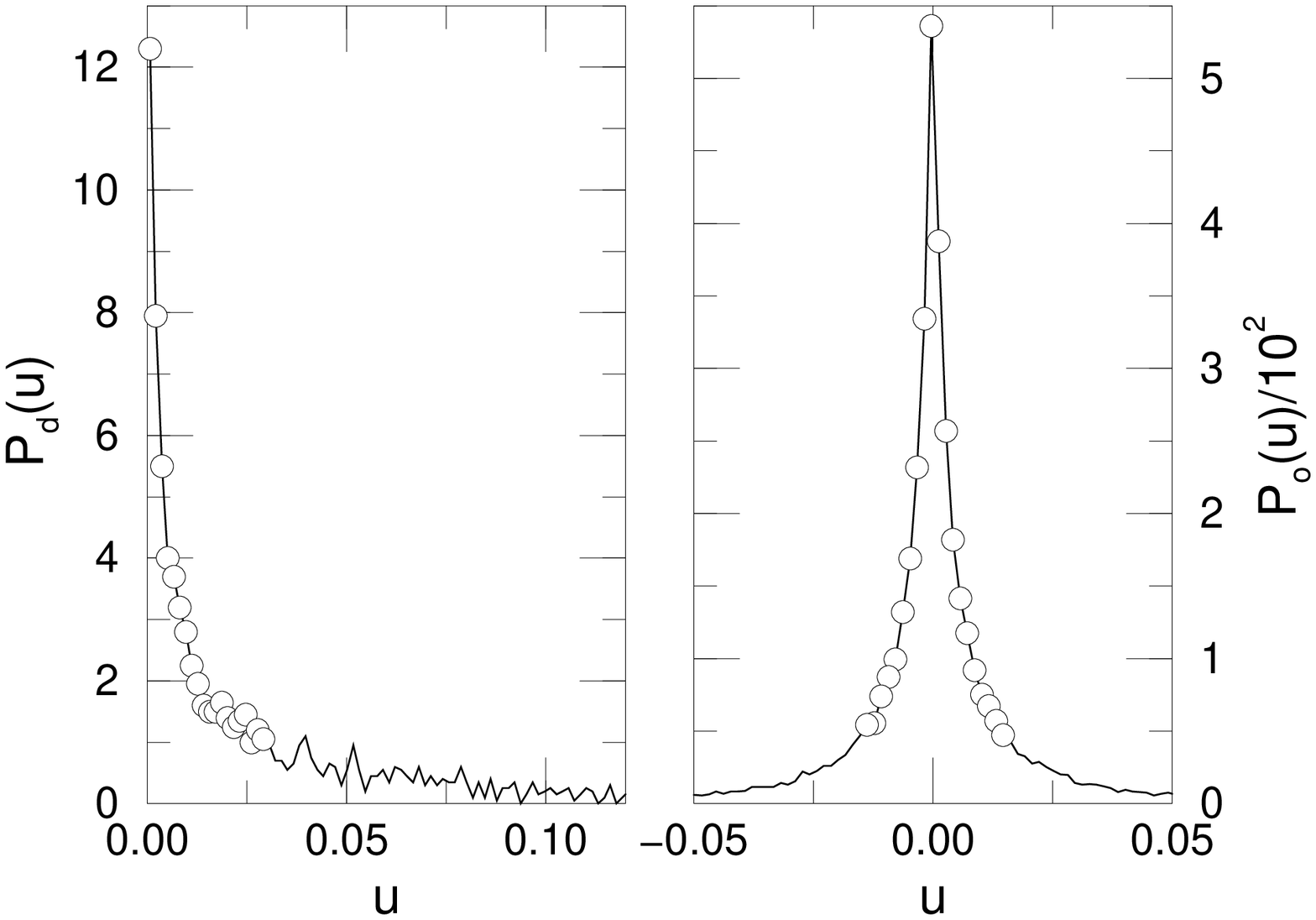}}
   \caption{
     Unnormalized distribution for the diagonal (left panel) and
     off-diagonal (right panel) coupling matrix elements $u$ with bin
     width $\Delta= 0.0015$ for the perturbed 2D Anderson model with
     $\lambda_1=3.1$ ($W=12$) and $M=25$. The circles indicate the 20
     smallest $u$ (largest $P_{\text{o}}(u)$) for diagonal
     (off-diagonal) data.}
   \label{fig-d2me-bin}
\end{figure}

\begin{figure} 
   \epsfxsize=0.8\columnwidth
   \centerline{\epsffile{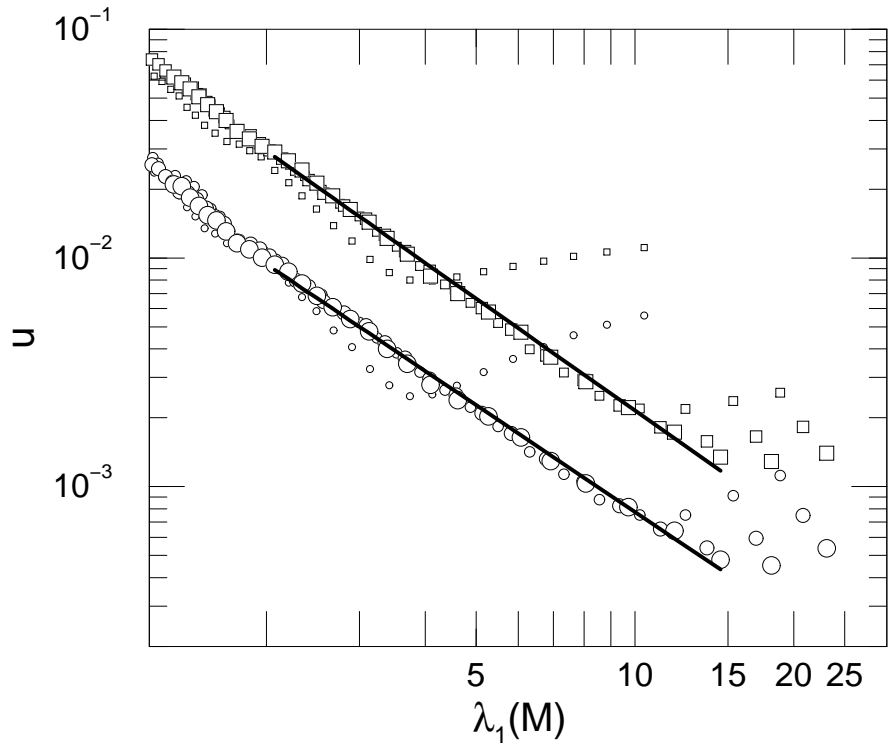}}
   \caption{
     Dependence of $u_{\text{abs}}$ (squares) and $u_{\text{typ}}$
     (circles) on $\lambda_1(M)$ for the perturbed 2D Anderson model and
     $M= 10, 25, 30$ and $35$ indicated by increasing symbol size. The
     solid lines represent the power laws $u_{\text{abs}} \sim
     \lambda_1^{-1.6}$ and $u_{\text{typ}} \sim \lambda_1^{-1.5}$. }
   \label{fig-d2me-l1}
\end{figure}

In Fig.\ \ref{fig-d2me-bin} we show the computed distributions
$P_{\text{d/o}}(u)$ for the present model. As for the TIP model the
diagonal elements are non-negative and $P_{\text{d}}(u)$ has a large
peak at $u=0$; $P_{\text{o}}(u)$ is again strongly non-Gaussian. The
results for $u_{\text{abs}}$ and $u_{\text{typ}}$ are presented in
Fig.\ \ref{fig-d2me-l1}. The dependence of $u_{\text{abs}}$ on
$\lambda_1(M)$ for $2 \leq \lambda_1(M) \leq 12$ follows
$u_{\text{abs}} \propto \lambda_1(M)^{-1.6 \pm 0.1}$ in agreement with
our above prediction.  Furthermore, here we also have $u_{\text{typ}}
\propto \lambda_1(M)^{-1.5 \pm 0.1}$.  As before, we note that the
slopes of $u_{\text{abs}}$ and $u_{\text{typ}}$ become smaller for
$\lambda_1(M)\approx M/2$ due to the finite sample sizes.  This
finite-size effect is just the same as for TIP and thus further
supports our use of the finite-size values $\lambda_1(M)$. We remark
that if instead of $\lambda_1(M)$, we use $\lambda_1(50)$ for plotting
the $u_{\text{abs}}$ and $u_{\text{typ}}$ data, that is irrespective
of the system sizes for which they had been computed, we obtain
$u_{\text{abs}} \propto \lambda_1^{-1.54 \pm 0.10}$ and
$u_{\text{typ}} \propto \lambda_1^{-1.47 \pm 0.10}$.  Thus both
choices of $\lambda_1$ show that $u_{\text{abs}}$ and $u_{\text{typ}}$
vary as $\lambda_1^{-1.5}$ within the accuracy of the calculation.

Since our toy model is mapped onto the same RMM as the TIP problem the
resulting localization length along the diagonal is also given by Eq.\ 
(\ref{eq-shep}). We thus arrive at the surprising conclusion, that
adding a weak random potential at the diagonal of a 2D Anderson model
leads to an enormous enhancement of the localization length along this
diagonal, in contradiction to the expectation expressed above, viz.\
that increasing disorder leads to stronger localization.


\subsection{1D Anderson model with perturbation}

An even more striking contradiction can be obtained for a 1D Anderson
model of localization. The eigenstates are again given by Eq.\ 
(\ref{eq-psi1}) with $\lambda_1$ known from second order perturbation
theory \cite{thouless} and numerical calculations \cite{numtmm} to
vary as $\lambda_1 \sim V^2/W^2$ for small disorder.  We now add a
weak random potential of strength $U$ at all sites. Since the result
is obviously a 1D Anderson model with a slightly higher disorder
strength the localization length will be reduced, $\lambda_1(U) \sim
V^2/(W^2+U^2)$.
Now we map onto an RMM according to Refs.\ \onlinecite{shep,shep2}. The
additional potential leads to transitions between the unperturbed
eigenstates $\psi_n$. Each such state is now coupled to ${\cal
  O}(\lambda_1)$ other states by coupling matrix elements $\langle
\psi_{n}|U|\psi_{n'} \rangle$ with magnitude $u \sim U
\lambda_1^{-1/2}$ since we sum over $\lambda_1$ contributions with
magnitude $U/\lambda_1$ and supposedly random phases.

\begin{figure} 
   \epsfxsize=0.8\columnwidth
   \centerline{\epsffile{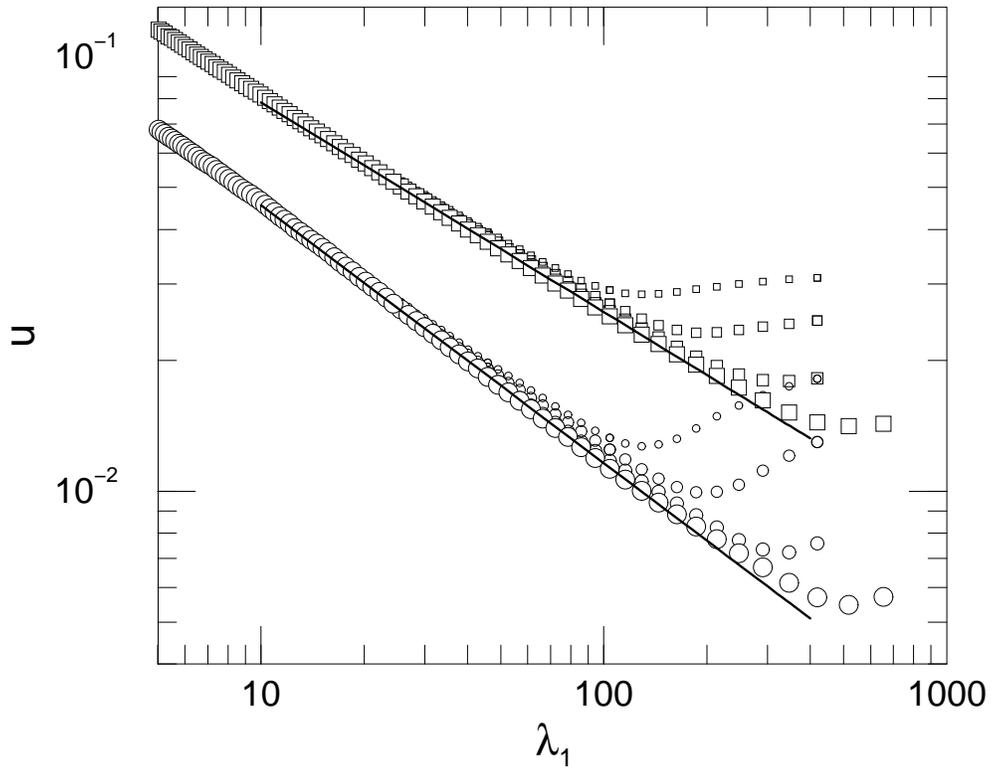}}
   \caption{
     Dependence of $u_{\text{abs}}$ (squares) and $u_{\text{typ}}$
     (circles) on $\lambda_1$ for the perturbed 1D Anderson model and
     $M= 200, 300, 500$ and $800$ indicated by increasing symbol
     size. The solid lines represent the power laws $u_{\text{abs}}
     \sim \lambda_1^{-0.48}$ and $u_{\text{typ}} \sim
     \lambda_1^{-0.59}$. }
   \label{fig-d1me-l1}
\end{figure}

\begin{figure} 
   \epsfxsize=0.8\columnwidth
   \centerline{\epsffile{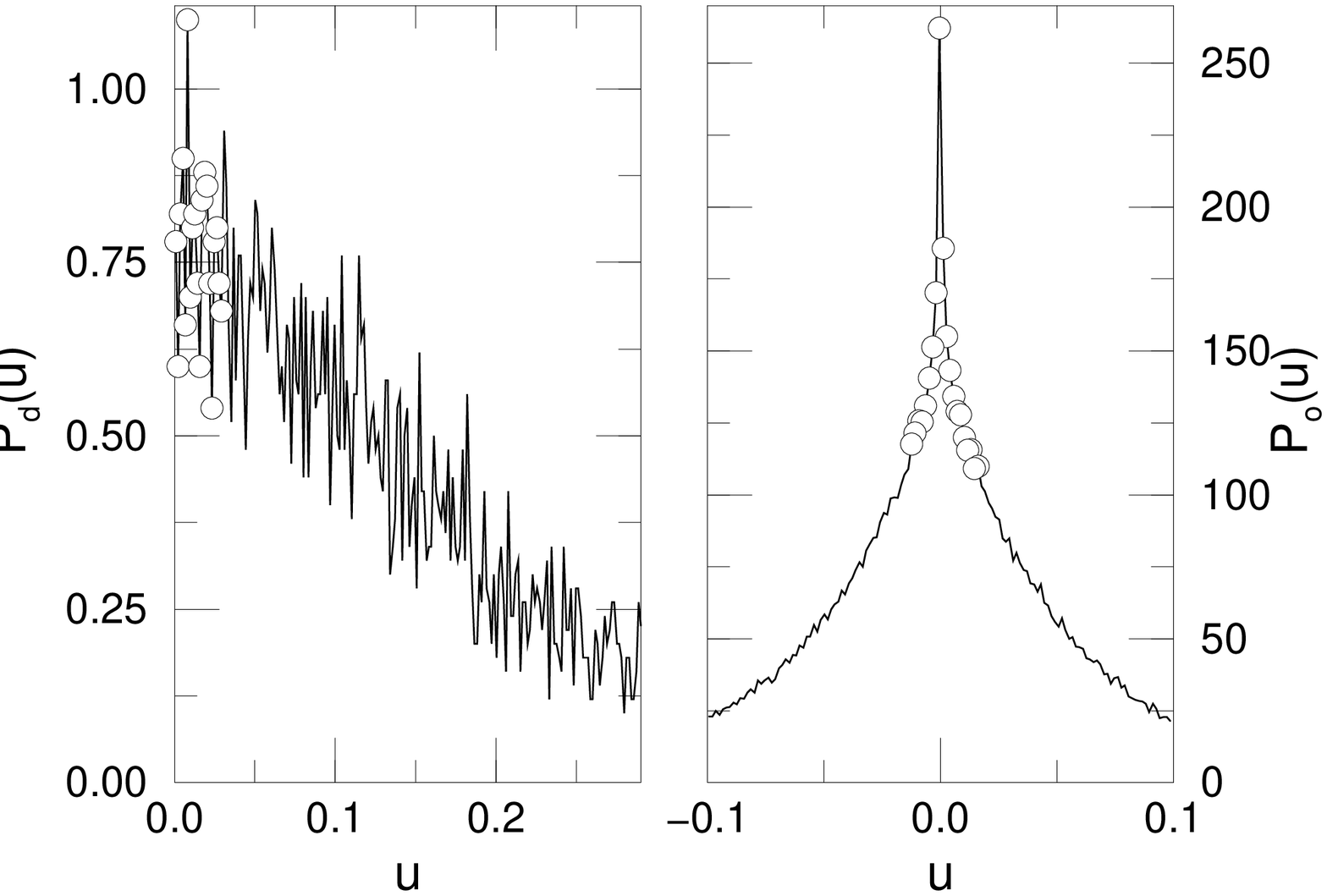}}
   \caption{
     Unnormalized distribution for the diagonal (left panel) and
     off-diagonal (right panel) coupling matrix elements $u$ with bin
     width $\Delta= 0.0015$ for the perturbed 1D Anderson model with
     $\lambda_1=26$ ($W=2$) and $M=200$. The circles indicate the 20
     smallest $u$ (largest $P_{\text{o}}(u)$) for the diagonal
     (off-diagonal) data.}
   \label{fig-d1me-bin}
\end{figure}

Again we numerically check the relation between $u_{\text{abs}}$ and
$u_{\text{typ}}$ as functions of $\lambda_1$. In Fig.\ 
\ref{fig-d1me-l1}, we show results obtained for chains with various
lengths and $50$ disorder configurations for each $W$.  $\lambda_1$ is
computed by TMM as in section \ref{sec-num}.  In Fig.\ 
\ref{fig-d1me-bin} we show the distributions $P_{\text{d/o}}(u)$. We
note that $P_{\text{o}}(u)$ is non-Gaussian as for the TIP model and
the perturbed 2D Anderson model. $P_{\text{d}}(u)$ is similar to the
previous models, but the fluctuations are much larger. For $10 \leq
\lambda_1 \leq 250$, $u_{\text{abs}}$ varies as $\lambda_1^{-0.48 \pm
  0.10}$ as we predicted above. $u_{\text{typ}}$ varies as
$\lambda_1^{-0.59 \pm 0.10}$. Both variations are compatible with
$\alpha= 1/2$.  Again we need at least $\lambda_1 \gtrsim M/2$ in
order to suppress the effects of the finite chain lengths.

In analogy to section \ref{sec-rmm} the application of Fermi's golden
rule in this 1D case leads to a diffusion constant $D \sim U^2
\lambda_1^2 /V$. The number of states visited within a time $\tau$ is
now $N \sim U \lambda_1 V^{-1/2} \tau^{1/2}$.  Again, diffusion stops
at a time $\tau^*$ when the level spacing of the states visited equals
the frequency resolution.  This gives $\tau^* \sim U^2 \lambda_1^2
/V^3$.  The localization length $\lambda$ of the perturbed system thus
reads $\lambda \sim \sqrt{D \tau^*} \sim U^2 \lambda_1^2$ as in Eq.\ 
(\ref{eq-shep}), in clear contradiction to the correct result.
%

%
%

\section{Failure of the BSP for toy models}
\label{sec-bsptoy}

We now discuss the relation of our results to Imry's BSP \cite{imry}
for the TIP problem.  In this approach one considers blocks of linear
size $\lambda_1$ and calculates the dimensionless pair conductance on
that scale,
\begin{equation}
  g_2 \sim { u^2 \over \Delta^2},
  \label{eq-bsp}
\end{equation}
where $u$ represents the typical interaction-induced coupling matrix
element between states in neighboring blocks and $\Delta \sim
V/\lambda_1^2$ is the level spacing within the block. If the typical
coupling matrix element depends on $\lambda_1$ as $u \sim U
\lambda_1^{-\alpha}$ the pair conductance obeys
\begin{equation}
  g_2 \sim  (U/V)^2 \lambda_1^{4-2\alpha}.
\end{equation}
Again, an estimate analogous to Shepelyansky's (\ref{eq-matrix}) gives
$\alpha=3/2$ which leads to a strong enhancement of the pair
conductance $g_2 \sim \lambda_1$ as compared to the SP conductance
$g_1$ which is of order unity on scale $\lambda_1$. In contrast, the
numerical data of section \ref{sec-num} suggest that the pair
conductance increases much less, viz.\ $g_2 \sim (U/V)^2
\lambda_1^{0.1\pm 0.2}$ for the fitted exponent $\alpha=1.95\pm 0.10$.
Asymptotically for large $\lambda_1$ the pair conductance is likely to
be enhanced even less than that. The behavior will be close to or even
smaller than the marginal case $g_2 \sim g_1$.  All this is in
complete agreement with our corresponding considerations for the RMM.

For the 2D Anderson model considered in the last section, the BSP can
be applied analoguously. Again, we consider blocks of linear size
$\lambda_1$ and compute the typical perturbation-induced matrix
elements between these blocks as in section \ref{sec-2dam}. We then
find that according to the BSP the conductance of a 2D Anderson model
with additional weak perturbing potential along the diagonal is given
by Eq.\ (\ref{eq-bsp}). Using $\alpha= 1.5 \pm 0.1$ as obtained in
section \ref{sec-2dam} from the numerical data for $u_{\text{abs}}$
and $u_{\text{typ}}$, we then have $g_2 \sim (U/V)^2 \lambda_1$. Thus
the BSP yields the same unphysical result as the RMM approach of
section \ref{sec-2dam}.

Let us also apply the BSP to the 1D toy example. The level spacing in
a 1D block of size $\lambda_1$ is $\Delta \sim V/\lambda_1$, and the
coupling matrix element between states in neighboring blocks is $t
\sim U \lambda_1^{-1/2}$.  Thus, the conductance of the perturbed
system on a scale $\lambda_1$ is obtained as $ g_p \sim (U/V)^2
\lambda_1$.  For large $\lambda_1$ this again contradicts the correct
result, viz.\ a decrease of the conductance compared to the
unperturbed system.

Thus, the BSP applied to the two toy models introduced in section
\ref{sec-toy} gives the same qualitatively incorrect results for the
localization properties as the RMM. This is not surprising since the
only ingredients of the BSP are the intra-block level spacing $\Delta
\sim V/\lambda_1^2$ and the inter-block coupling matrix elements $u$
which also enter the RMM and have been discussed in section
\ref{sec-toy}.

%
%

\section{Conclusions}
\label{sec-con}

To summarize, we have reinvestigated the RMM approach to the problem
of TIP in a random potential. We have shown that this kind of mapping an
interacting disordered system onto an effective random matrix model is
potentially dangerous since (A) it may overestimate the typical coupling 
matrix element and (B) it neglects correlations between the matrix elements.

In the first part of the paper we investigated the dependence of the
matrix elements entering the RMM on the SP localization length
$\lambda_1$.  We found the dependence of the typical matrix element
$u_{\text{typ}}$ to be significantly stronger than for the averaged
absolute value $u_{\text{abs}}$ which is used in Refs.\ 
\onlinecite{shep,shep2,imry,weinmann,frahm3}. If the RMM approach of
section \ref{sec-rmm} is modified by using the numerically determined
relation between $u_{\text{typ}}$ and $\lambda_1$ instead of Eq.\ 
(\ref{eq-matrix}) the resulting enhancement of $\lambda_2$ with
respect to $\lambda_1$ becomes much weaker. We showed that the
difference between $u_{\text{typ}}$ and $u_{\text{abs}}$ is due to the
over-simplified assumption (A) that the wave functions behave
according to Eq.\ (\ref{eq-psi1}). Moreover, our data for
$u_{\text{typ}}$ show systematic deviations from power-law behavior
indicating that the true asymptotic dependence of $u_{\text{typ}}$ on
$\lambda_1$ is likely to be very close to or stronger than the
marginal case $u_{\text{typ}} \sim \lambda_1^{-2}$. If the asymptotic
dependence is stronger than $u_{\text{typ}} \sim \lambda_1^{-2}$ the
Shepelyansky enhancement vanishes in the limit of large $\lambda_1$
even within the RMM approach.

In the second part of this paper we showed that there are physical
situations where mapping onto an RMM as in Ref.\ \onlinecite{shep}
gives qualitatively incorrect results, {\em e.g.}, an increase of the
localization length in physical situations where it should rather
decrease. This failure occurs even if the RMM contains the correct
dependence of $u_{\text{typ}}$ on $\lambda_1$. This shows in contrast
to assumption (B) that in general the correlations between the matrix
elements cannot be neglected since they contain information essential
for the interference leading to Anderson localization.  Note that the
approach of Ref.\ \onlinecite{PS}, while correcting assumption (A),
still includes a mapping onto an RMM and thus is plagued by the same
problems as assumption (B).  Analogously, in Ref.
\onlinecite{jacquod2} the decay rate $\Gamma$ is calculated
numerically, avoiding assumption (A). However, the formula
$\lambda_2/\lambda_1 \sim \Gamma \lambda_1^2/V$ employed in Ref.\ 
\onlinecite{jacquod2} is also based on an assumption similar to (B).

Let us comment on the relevance of this work for the original problem
of TIP in a random potential. None of our results constitute, of
course, a proof that the enhancement of the TIP localization length
$\lambda_2$ predicted in Ref.\ \onlinecite{shep} does not exist.
However, in our opinion, the toy counter examples to the RMM approach
introduced in section \ref{sec-toy} let the analytical arguments
giving Eq.\ (\ref{eq-shep}) appear much weaker. Taking the RMM
approach seriously but using the numerical results for the typical
coupling matrix element presented in section \ref{sec-num} we find
that the dependence of the enhancement factor $\lambda_2/\lambda_1$ on
$\lambda_1$ is significantly weaker than in Eq.\ (\ref{eq-shep}).
Nevertheless, an enhancement of the pair localization length for TIP
as compared to $\lambda_1$ may still exist, although the underlying
mechanism should then be different.  Results supporting such an
enhancement have been obtained by Green function methods \cite{oppen}
together with finite-size scaling arguments.\cite{kim} The most recent
data obtained in Ref.\ \onlinecite{kim} finds an exponent $\alpha =
1.45 \pm 0.2$ for $U=1$.

\acknowledgments

We thank Frank Milde for programming help and discussions.  This work
was supported by the Deutsche Forschungsgemeinschaft through grants
Vo659/1, Schr231/13, SFB 393 and by the National Science Foundation
through grant DMR-95-10185.

%
%

%
%
\newpage

\end{document}